\documentclass[runningheads]{llncs}
\usepackage{authblk}
\usepackage[T1]{fontenc}
\usepackage{textcomp}
\usepackage{multirow}
\usepackage{hyperref}
\usepackage{amsmath,amssymb}
\usepackage{float}
\usepackage{pgfplots} 
\newcommand{\mathz}{\ooalign{$z$\cr\hfil\rule[.5ex]{.2em}{.06ex}\hfil\cr}}

\usepackage{graphicx}
\usepackage{subcaption}
\begin{document}
\title{LRH-Net: A Multi-Level Knowledge Distillation Approach for Low-Resource Heart Network}
\titlerunning{LRH-Net}
\author{Ekansh Chauhan\inst{1}\and
Swathi Guptha\inst{1} \and
Likith Reddy \inst{1} \and
Bapi Raju \inst{1}}
\authorrunning{E. Chauhan et al.}
\institute{International Institute of Information Technology, Hyderabad, India \email{ekansh.chauhan@research.iiit.ac.in}}
\maketitle              % typeset the header of the contribution
\begin{abstract}

An electrocardiogram (ECG) monitors the electrical activity generated by the heart and is used to detect fatal cardiovascular diseases (CVDs). Conventionally, to capture the precise electrical activity, clinical experts use multiple-lead ECGs (typically 12 leads). Recently, large-scale deep learning models have been used to detect these diseases, however, they require large memory and long inference time. We propose a low-parameter model, Low Resource Heart-Network (LRH-Net), that detects ECG anomalies in a resource-constrained environment. On top, multi-level knowledge distillation (MLKD) is employed to improve model generalization. MLKD distils the dark-knowledge from higher parameter models (teachers) trained on different lead configurations to LRH-Net. The LRH-Net has 106x fewer parameters and 76\% faster inference than the teacher model for detecting CVDs. Using MLKD, the performance of LRH-Net on reduced lead data was scaled up to 3.25\%, making it suitable for edge devices.


\end{abstract}
\section{Introduction}
One of the most common causes of death around the globe is cardiovascular diseases (CVDs). According to WHO, in 2021, around 32\% of all deaths, i.e., 17.9 million people died from CVDs~\cite{1}. These diseases manifest with no severe symptoms and are difficult to diagnose, leading to underestimating the risk or severity. Thus, early diagnosis of these diseases can potentially save millions of lives~\cite{2}. 

Electrocardiography (ECG) is a low-cost and widely used process to monitor abnormal electrical activity in the heart~\cite{3}. However, this process can only be used and interpreted by a cardiologist~\cite{4}. The advancement of the Internet of Things (IoT) makes real-time capturing of ECG signals feasible using wearable devices. Thereby resulting in massive ECG data, which is used in machine learning techniques to detect CVDs~\cite{6,32}.

Most of the early literature on CVD used classical feature extraction approaches along with machine learning models~\cite{13,15,16,17}. Then, artificial neural networks such as multilayer perceptrons demonstrated great performance and removed the requirement of manually handpicked features to some extent, especially with the advent of deep neural networks~\cite{18,19}. However, these networks are typically unconcerned with power consumption, memory consumption and execution time, preferring to be more accurate~\cite{24}, making them difficult for deploying on low-compute resources. There has always been a trade-off between performance and size -- trading off the extent to which size should be decreased and yet retain acceptable performance.

In recent years, most of the work is pivoted on capturing ECG signals using wearable devices using Bluetooth and Internet connectivity of the mobile phones which are later processed on cloud to detect the anomalies. Furthermore, traditional electrocardiography setups use 12 electrodes to monitor heart activity, but using such a setup in a real-time environment would require excessive computation and be inconvenient or tedious process for the end user. Rural areas, on the other hand, have significant contributions to cardio-vascular disease burden, and finding such compute resources is difficult there~\cite{1,7}. Therefore, an efficient neural network which takes data from fewer electrodes and requires less memory and inference time is required for an edge computing wearable device.

To the best of our knowledge there is no other solution proposed in the literature for resource constrained environments while considering heterogeneity in datasets (and disease conditions). We propose a low-parameter model called Low Resource Heart-Network (\emph{LRH-Net}) on top of which Multi-Level Knowledge Distillation (MLKD) methods are also proposed to enhance its performance. This novel approach is compared with an existing high-performance large-scale model~\cite{26} and a commonly used low-scale model~\cite{20} baselines on heterogeneous dataset~\cite{Dataset}. The source code for the proposed model and all the experiments that are done are made public to motivate further research in this field~\footnote{https://github.com/ekansh09/LRH-Net}. 

Main contributions of this paper are:
\begin{enumerate}
\item	A real-time cardiovascular disease detection model which is 106x smaller than a large-scale model and 12x times smaller than the existing low-scale model.
\item	A Multi-Level knowledge distillation approach to improve the performance of LRH-Net (student model) and to reduce the number of electrodes and input leads data required for the student model .
\item	Performed evaluation on a very diverse, publicly available and combination of multiple datasets to increase its desirability.
\end{enumerate}

\section{Methodology}

\subsection{Pre-Processing}
The sampling frequency ranges from 257Hz to 1KHz in the datasets being used. As part of pre-processing, we resampled the data to 257Hz, the minimum in our case. Each ECG is set to be 4096 points long, approximately 16 seconds. The time series is randomly clipped for longer duration and zero-padded for shorter duration signals in order to give a fixed sequence length as an input to the deep learning models. The signal is then normalized using z-score to remove technological biases between datasets i.e., a signal $x_n \in n^{th}$  channel (lead) was transformed using Equation \ref{eq:1}, where $\overline{x_n}$ is the mean and $\sigma_n$ is the standard deviation across the $n^{th}$ channel.
\begin{equation}\label{eq:1}
 x_n =  \frac {x_n - \overline{x_n}}{\sigma_n}
\end{equation}

Finally, we also took one-hot encoded phenotypic information such as age (scaled between 0 and 1) and gender, into consideration and represented missing values with additional two mask variables.

\subsection{Architecture}
\begin{figure}
\centering
\includegraphics[width=0.7\textwidth]{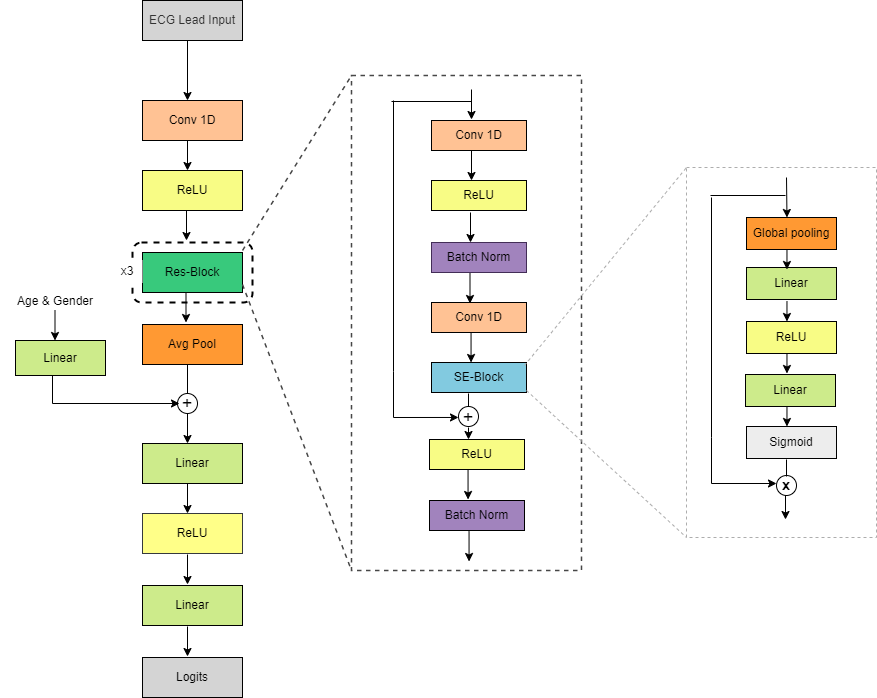}
\caption{LRH-Net architecture: Proposed low parametric Model} \label{fig1}
\end{figure}

The proposed model, LRH-Net, inspired from the ResNet architecture~\cite{25} with 3 residual blocks (Res-Blocks) in it and is depicted in Fig.~\ref{fig1}. The motivation for using a Resnet based architecture is the power of skip connections that ameliorate the vanishing gradients problem of the back propagation learning scheme and enhance model convergence. Each residual block has two convolution layers, ReLU activation function, batch normalization and one squeeze and excitation (SE) block~\cite{33}. The starting filter is always 16 and increased by a factor of two in the case of Res-blocks. SE-Block aids in learning the importance of various features and paying more attention to those that are more important, thereby improving classification performance. We used it to model the spatial relationship among the ECG channels. Additionally,~\cite{26} showed that integrating patient's age and gender improves the performance and is easy to feed into an edge device. Considering this, we passed these values through a linear layer followed by concatenation to the features obtained from the average pool layer, that then passes all of them through two more linear layers with ReLU in between in the model to generate the logits.

Knowledge distillation (KD) refers to the idea of model compression where the small model (student model) mimics the larger model (teacher model) using soft labels provided by the teacher model~\cite{10}. We used this knowledge distillation method to further improve the performance of LRH-Net. While distilling information from a large network to a smaller network, it is preferable to use a similar kind of architecture for the distillation training~\cite{11}. Hence, we used the runner-up network (SE-Resnet) of PhysioNet-2020 challenge proposed in~\cite{26} as our Teacher Network ($\Theta_{T}$).

\begin{figure}
     \begin{subfigure}[b]{0.5\textwidth}
         \includegraphics[width=0.9\textwidth]{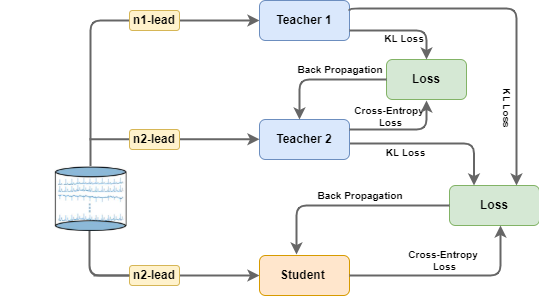}
         \caption{Sequential Configuration (s-MLKD)}
         \label{fig:Sequential MLKD}
     \end{subfigure}
     \hfill
     \begin{subfigure}[b]{0.5\textwidth}
         \includegraphics[width=0.9\textwidth]{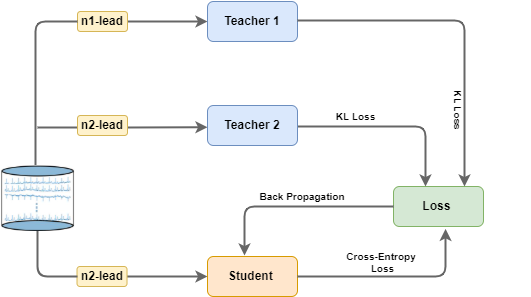}
         \caption{Parallel Configuration (p-MLKD)}
         \label{fig:Parallel MLKD}
     \end{subfigure}
        \caption{Multi-Level Knowledge Distillation (MLKD) Methods.}
        \label{MLKD}
\end{figure}

Recent work on knowledge distillation methods led to an idea of having a multi-teacher approach to reduce the gap between the high parameter model and the low parameter target model by introducing an intermediate size parameter model~\cite{11}. Using this as inspiration, we propose a Multi-Level Knowledge Distillation (MLKD) approach to reduce the number of electrodes required to generate fewer lead data and simultaneously enhance the performance of LRH-Net in multiple steps, in a sequential or parallel configuration, to retain the knowledge (representations) from a large-scale model trained on multi-lead ECG data (see Fig.~\ref{MLKD}). Both sequential and parallel configurations are put to the test in a 2-step procedure. First, by decreasing the number of input channels while maintaining the network’s size, and secondly, by reducing the network’s size while keeping the input channels constant.

Let $n$ be the number of channels (leads). $\Theta_{T}^{n_1}$ be the pre-trained teacher network with $n_1$ input channels and $\Theta_{T}^{n_2}$ be the teacher network with $n_2~(n_2<n_1)$ input channels. We use Binary cross-entropy (BCE) loss (Equation~\ref{eq:2}) as student loss and Kullback–Leibler divergence (KL) loss (Equation~\ref{eq:3}) as distillation loss to account the knowledge transfer from the teacher model to the student model.

\subsubsection{Sequential MLKD (s-MLKD)}: At step one ($t=1$), knowledge is distilled (KD) from $\Theta_{T}^{n_1}$ to $\Theta_{T}^{n_2}$. Therefore, from Equations~\ref{eq:4} and~\ref{eq:5}, the loss at this step is given by $\mathcal{L}(\Theta_{T}^{n_2}, \Theta_{T}^{n_1})$. Now, let the distilled $\Theta_{T}^{n_2}$ be $\Theta_{T_d}^{n_2}$ , where $d$ denotes the distilled model. At second step, we perform KD from $\Theta_{T}^{n_1}$ and $\Theta_{T_d}^{n_2}$ to LRH-Net with $n_2$ input channels ($\Theta_{L}^{n_2}$). We give different weightage to knowledge obtained from each teacher model, denoted as $\beta$ in Equation \ref{eq:4}. Therefore, the final loss function for a 2-step s-MLKD method is given at 2nd-step as  $\mathcal{L}(\Theta_{L}^{n_2}, \Theta_{T_d}^{n_2}, \Theta_{T}^{n_1})$. Here, steps are equal to the number of teacher networks, i.e. after each step (except last), we get a trained intermediate teacher network.

\begin{equation}\label{eq:2}
\resizebox{.70\hsize}{!}{$BCE\left(p,y\right)=  \frac{-1}{N} \sum_{i}^{N}\sum_{j}^{M} y_{ij} \log p_{ij} + \left(1-y_{ij}\right) \log \left(1-p_{ij}\right)$}
\end{equation}

\begin{equation}\label{eq:3}
\resizebox{.36\hsize}{!}{$ KL\left(p,y\right)= \sum_{x \in X} y\left(x\right) \log \frac{y\left(x\right)}{p\left(x\right)}$}
\end{equation}

% \begin{equation}\label{eq:4}
% \omega_s^{\left(A_1,A_2,\ldots ,A_{s+1}\right)} = \alpha \cdot BCE\left(P_1\left(A_1^{logits}\right), GT\right) + \left(1-\alpha\right)\cdot\sum_{x \in s} \beta_x \cdot KL\left(\frac{\log\left(P_2\left(A_1^{logits}\right)\right)}{T}, \frac{P_2\left (A_{x+1}^{logits}\right)}{T}\right)
% \end{equation}

\begin{equation}\label{eq:4}
\resizebox{.75\hsize}{!}{$ \mbox{\emph{MLKD}}_{loss}\left(\Theta_0,\Theta_1,\ldots,\Theta_{t}\right) = \sum_{x=1}^{t} \beta_x KL\left(\frac{\log\left(s\left(\mathz_0\right)\right)}{\tau}, \frac{s\left (\mathz_{x}\right)}{\tau}\right)$}
\end{equation}

\begin{equation}\label{eq:5}
\resizebox{.92\hsize}{!}{$\mathcal{L}{\left(\Theta_0,\Theta_1,\ldots,\Theta_{t} ; y\right)}  =  \lambda  BCE\left(\sigma\left(\mathz_{\Theta_0}\right), y\right) + \left(1-\lambda\right) \mbox{\emph{MLKD}}_{loss}\left(\Theta_0,\Theta_1,\ldots,\Theta_{t}\right)$}
\end{equation}

Where, $p,y$ are probability and ground truth values. In Equation \ref{eq:2}, $N$ is number of samples in each batch and M is number of classes. In Equation \ref{eq:3}, $X$ is probability space of $p,y$. Equation \ref{eq:5} represents the loss at every step and in the series notation $(\Theta_0,\Theta_1,\ldots,\Theta_{t})$, $\Theta_0$ is student and all other are teacher networks. $\lambda$ parameter implements the trade-off between BCE-loss and MLKD-loss. $\mathz_{t}$ represents the logits of $\Theta_t$ network. $\sigma$ is the \emph{Sigmoid} activation function, $s$ is the \emph{SoftMax} activation function and $\tau$ is the temperature hyper-parameter used to generate soft-labels. The log-softmax is applied to student's logits in MLKD loss for numerical stability in Pytorch. 

\subsubsection{Parallel MLKD (p-MLKD)}: In this scheme, the first step is to train $\Theta_{T}^{n_2}$ independently and the second step looks almost like the s-MLKD scheme, $\Theta_{T}^{n_2}$ being different. In p-MLKD, we use two independently pre-trained networks to teach a student network. So, from Equations~\ref{eq:4} \&~\ref{eq:5}, the final loss function for p-MKLD will be  $\mathcal{L}(\Theta_{L}^{n_2}, \Theta_{T}^{n_2}, \Theta_{T}^{n_1})$.

Finally, with s- and p-MLKD schemes, the distilled LRH-Net has the dark knowledge of 12 leads but takes fewer leads as input and outputs logits from the last dense layer. Probability scores are obtained by applying $\sigma$ (sigmoid) to the output from logits block. Then, a differential evolution genetic technique is used to optimize class thresholds~\cite{27}. Our experiments reveal that these thresholds do not vary as the number of leads are varied.

\section{Experiments}
\subsection{Dataset}
\begin{figure}
\centering
\includegraphics[width=\textwidth]{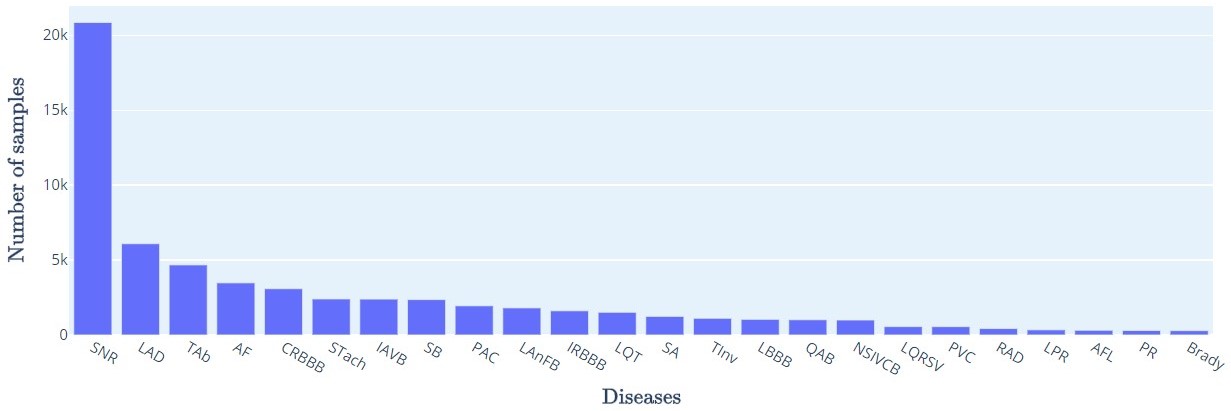}
\caption{Dataset Distribution} \label{Dataset}
\end{figure}

A total of 43101 standard 12-lead ECG (I, II, III, aVL, aVR, aVF, V1-V6) recordings are used from four publicly available datasets provided by Physionet-2020 challenge~\cite{Dataset}, i.e., CPSC Database and CPSC-Extra Database, INCART Database, PTB and PTB-XL Database, and the Georgia 12-lead ECG Challenge. It has 24 unique class labels and a signal may have more than one class label assigned to it. The distribution among these 24 classes in the dataset can be visualized from Fig~\ref{Dataset}. The signal length varies from 10 seconds to 30 minutes.

% \begin{table}[h]
% \caption{Dataset Description}
% \label{tab:Dataset_des}
% \begin{tabular}{|c |c |c |}
% \hline
% \textbf{S.No.} & \textbf{Disease} & \textbf{Total} \\ \hline
% 1 & Bradycardia   (Brady) & 288 \\ \hline
% 2 & Pacing   rhythm (PR) & 299 \\ \hline
% 3 & Atrial   flutter (AFL) & 314 \\ \hline
% 4 & Prolonged   pr interval (LPR) & 340 \\ \hline
% 5 & Right   axis deviation (RAD) & 427 \\ \hline
% 6 & Premature   ventricular contractions (PVC) & 551 \\ \hline
% 7 & Low   qrs voltages (LQRSV) & 556 \\ \hline
% 8 & Nonspecific   intraventricular conduction disorder (NSIVCB) & 997 \\ \hline
% 9 & qwave   abnormal (QAB) & 1013 \\ \hline
% 10 & Left   bundle branch block (LBBB) & 1041 \\ \hline
% 11 & t   wave inversion (TInv) & 1112 \\ \hline
% 12 & Sinus   arrhythmia (SA) & 1240 \\ \hline
% 13 & Prolonged   qt interval (LQT) & 1513 \\ \hline
% 14 & Incomplete   right bundle branch block (IRBBB) & 1611 \\ \hline
% 15 & Left   anterior fascicular block (LAnFB) & 1806 \\ \hline
% 16 & Premature   atrial contraction (PAC) & 1942 \\ \hline
% 17 & Sinus   bradycardia (SB) & 2359 \\ \hline
% 18 & 1st   degree av block (IAVB) & 2394 \\ \hline
% 19 & Sinus   tachycardia (STach) & 2402 \\ \hline
% 20 & Complete   right bundle branch block (CRBBB) & 3071 \\ \hline
% 21 & Atrial   fibrillation (Af) & 3475 \\ \hline
% 22 & t   wave abnormal (TAb) & 4673 \\ \hline
% 23 & Left   axis deviation (LAD) & 6086 \\ \hline
% 24 & Sinus   rhythm (SNR) & 20846 \\ \hline
% \end{tabular}
% \end{table}

\subsection{Implementation Details}
The Pytorch framework is used to create the models. LRH-Net is trained for 1 hour 40 minutes using the Adam optimizer with L2 weight decay of 5e-4 for 90 epochs with a batch size of 64. We start with a learning rate of 0.001 and utilise the StepLr scheduler with step size of 20 to change it throughout the training. $\alpha$ and $\tau$ are set to 0.3, 7 respectively. Since we used a 2-step approach, the weight list given to the teachers, $\beta$, is $[0.4,0.6]$. It has 84,516 trainable parameters. All the experiments are carried out using Nvidia Tesla P100 GPUs.

\section{Results and Discussion}
5-Fold cross-validation is employed to evaluate LRH-Net using the metric provided by the Physionet-2020 challenge called challenge metric score (CM-Score) and the micro-F1-score. Misdiagnoses that result in treatments or outcomes that are similar to the true diagnosis as determined by the cardiologist are given partial credit in the challenge metric. It reflects the clinical reality that some misdiagnoses are more harmful than others and should be scored accordingly~\cite{Dataset}. The F1-score is also reported as it more accurately reflects the performance on an imbalanced class data set in one vs all setup. 

\subsection{Baselines}
There are no methods available to reduce ECG leads without compromising on the knowledge of all leads. Also, none of the previously available low-parameter models have trained on diverse multiple-datasets as provided in~\cite{Dataset}. Thus, we use four KD techniques mentioned below (see Table~\ref{tab3}) and the following two models as our baselines, SE-Resnet (our teacher) and 1D-CNN (see Table~\ref{tab1},~\ref{tab2}).

\begin{table}
\caption{Comparison of LRH-Net in terms of the number of trainable parameters, size and inference time with baseline models.}
\label{tab1}
\centering
\begin{tabular}{|l|l|l|l|}
\hline
Model & Parameters & Size & Inference Time \\
\hline
\emph{SE-Resnet} &  \emph{8.9M} & \emph{35.30 Mb} & \emph{3.43 seconds}\\
\hline
1D-CNN &  994K & 3.88 Mb & \textbf{0.47 seconds}\\
\textbf{LRH-Net} & \textbf{84K} & \textbf{0.35 Mb} & 0.84 seconds \\
\hline
\end{tabular}
\end{table}

\subsubsection{Parameters and Size:}
Table~\ref{tab1} shows comparison of our proposed model with both the baselines. Empirically, it is noted that the number of parameter are almost directly proportional to the size of the model. In comparison to the baselines, LRH-Net has the fewest parameters which results in a more compact and efficient network. Because LRH-Net's parameters are 106 times smaller than SE-Resnet and 12 times smaller than 1D-CNN, the model size (in mb) is likewise 101 times and 11 times lower respectively.

\subsubsection{Inference Time and Complexity:}
The inference time of a model is directly proportional to the model complexity. The inference time of LRH-Net is significantly smaller than SE-Resnet but slightly more than that of 1D-CNN due to the presence of additional squeeze-and-excitation block within the Res-block and larger kernel size. The kernel size is experimentally chosen and squeeze-and-excitation block is added to help the architecture to draw attention to the fact that the dataset has classes (cardiovascular anomalies) that are not spread out evenly~\cite{33}. 

LRH-Net makes a trade-off between memory consumption and complexity with
inference time, which results in superior performance when compared to the baseline (1D-CNN), which has less inference time but poor performance and high memory consumption.

Few studies~\cite{35,36} have shown that a 3-lead ECG (I, II, V2) contains the majority of the information found in a 12-lead ECG. Considering this, all the models are tested on standard 12-lead, 3-lead, and 2-lead configurations which are also provided by~\cite{34}. The 3-lead configuration is [I, II, V2] and for 2-lead, it is [I, II]. Information from 10, 5, and 4 electrodes (including ground electrode) are required to obtain the data for 12, 3, and 2 leads, respectively.

Table~\ref{tab2} shows the performance comparison between LRH-Net and baselines on various lead configurations. LRH-Net with 83,748 parameters outperforms the existing low-scale baseline model of 1D-CNN with 993,860 parameters by a significant margin for all the lead configurations, i.e., 3.07\%, 2.76\%, 1.41\% increment in CM scores for 12-lead, 3-lead, and 2-lead configurations, respectively. When LRH-Net is compared to SE-Resnet, the model size or parameters are drastically reduced (LRH-Net has fewer parameters), which results in a performance drop of about 10.5 percent (in CM scores) for all lead configurations.

\begin{table}
\caption{Comparison of LRH-Net in terms of performance with baseline models.}\label{tab2}
\centering
\begin{tabular}{|l|l|l|l||l|l|l|l|l|l|}
\hline
\multirow{2}{*}{Model}  &\multicolumn{3}{c|}{CM-Score} &\multicolumn{3}{c|}{F1-Score} \\
\cline{2-4}\cline{5-7}
& 12-Lead & 3-lead & 2-Lead & 12-Lead & 3-lead & 2-Lead
\\
\hline
\emph{SE-Resnet}& \emph{67.43} & \emph{65.37} & \emph{63.34} & \emph{76.65} & \emph{75.42} & \emph{74.86}\\
\hline
1D-CNN & 58.66 & 56.92 & 55.87 & 69.50 & 67.61 & 67.16\\
\textbf{LRH-Net} & \textbf{60.46} & \textbf{58.49} & \textbf{56.66} & \textbf{72.76} & \textbf{71.33} &\textbf{70.21} \\
\hline
\end{tabular}
\end{table} 

\subsubsection{Knowledge Distillation:}
The following four knowledge distillation techniques are tested to increase the performance of LRH-Net. Vanilla knowledge distillation~\cite{10} is a method of extracting dark knowledge from the logits of deep models. In Fitnets~\cite{28}, the main idea is to directly match the feature activation of the teacher and the student. In Cross Layer distillation~\cite{30}, each student layer distills knowledge contained in multiple layers rather than a single fixed intermediate layer from the teacher model. With progressive self-knowledge distillation (PS-KD)~\cite{29}, on the other hand, a student model itself becomes a teacher model.

%%%%%%%%%%%%%%%%%%%%%%%%%%%%%%%%%%%%%%%%%%%%%%%%%%%%%%%%%%%%%%%%%%%
\begin{table}
\caption{Evaluation of LRH-Net on various Knowledge Distillation techniques.}\label{tab3}
\centering
\begin{tabular}{|l|l|l|l||l|l|l|}
\hline
\multirow{3}{*}{KD Technique} & \multicolumn{6}{c|}{Distilled LR-HNet}\\
\cline{2-4} \cline{5-7}
& \multicolumn{3}{c|}{CM-Score} & \multicolumn{3}{c|}{F1-Score}
\\
\cline{2-4} \cline{5-7} 
& 12-Lead & 3-lead & 2-Lead & 12-Lead & 3-lead & 2-Lead
\\
\hline
\textbf{Vanilla KD} & \textbf{60.67} & \textbf{59.49} & \textbf{57.27} & \textbf{73.41} & 73.04 & \textbf{71.18}\\
FitNet &  60.46 & 58.88 & 57.20 & 72.22 & 72.61 & 70.22 \\
PS-KD & 57.81 & 55.80 & 55.58 & 68.62 & 68.05 & 68.03 \\
Cross-Layer & 60.17 & 59.04 & 57.13 & 73.34 & \textbf{73.28} & 70.69\\
\hline
\end{tabular}
\end{table}

Table~\ref{tab3} shows the performance of LRH-Net on multiple lead configurations after being distilled from the pre-trained SE-Resnet on 12-Lead input. When compared to all the recent knowledge distillation techniques, the vanilla KD delivers the best results for our use case. Surprisingly, LRH-Net performance deteriorated when combined with the newest KD-Methods, PS-KD and Cross-Layer and Fit-nets. Vanilla KD improved the CM-Score of non-distilled LRH-Net (as compared from LRH-Net results of Table~\ref{tab2}) by 0.35\%, 1.71\%, 1.08\% for 12-, 3- and 2-leads, respectively. Cross-Layer KD gives best F1-scores for 3-lead configuration. It can also be noted that the F1-Score is not directly proportional to CM-Score. This latter method is useful when each class is considered independently, i.e., assuming no correlation among the disease classes.

%%%%%%%%%%%%%%%%%%%%%%%%%%%%%%%%%%%%%%%%%%%%%%%%%%%%%%%%%%%%%%%%%%%
\subsubsection{Multi-Level Knowledge Distillation:}
To reduce the knowledge drop while distilling a low-lead, low-parameter model from a high-lead, high-parameter model, we use the multi-level knowledge distillation (MLKD) methods. These methods help the low-scale network in learning dark knowledge from the large-scale network in a step-by-step process. The CM-score and F1-scores of the LRH-Net with 3-lead and 2-lead inputs with sequential and parallel MLKD are reported in Table~\ref{tab4}. For both the configurations, the proposed MLKD incorporating Vanilla KD performs better than directly downgrading the leads using simply Vanilla KD (see results in Table~\ref{tab3}). The percentage increments in CM-Scores, when LRH-Net using MLKD and non-distilled LRH-Net (see Table~\ref{tab2}) are compared are 3.25\% and 3.12\% for 3- and 2-lead configurations, respectively.

\begin{table}
\caption{Evaluation of LRH-Net on MLKD methods with two teacher networks i.e. (step $(t) = 2$).}\label{tab4}
\centering
\begin{tabular}{|c|l|l|l|}
\hline
Number of Leads (n2) & KD-Technique & CM-Score & F1-Score \\
\hline
\multirow{2}{*}{3-Lead} & s-MLKD & \textbf{60.39} & \textbf{73.44}\\
 & p-MLKD & 60.36 & \textbf{72.51}\\
\hline
\multirow{2}{*}{2-Lead} & s-MLKD & 58.36 & 71.43\\
 & p-MLKD & \textbf{58.43} &  \textbf{72.66}\\
\hline
\end{tabular}
\end{table}

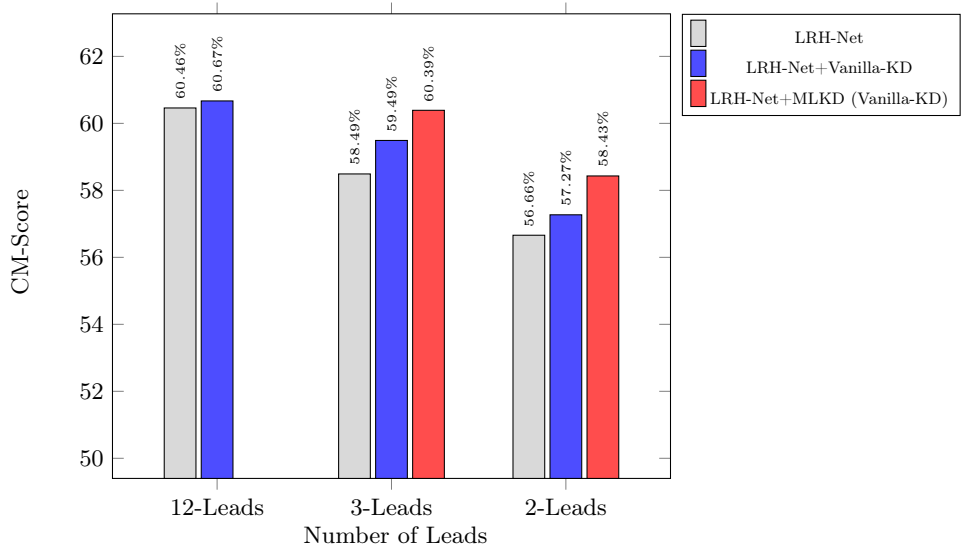
\begin{figure}
\centering
\begin{tikzpicture}  
\begin{axis}  
[  
    ybar, % ybar command displays the graph in horizontal form, while the xbar command displays the graph in vertical form. 
    bar width=12pt,
    ymin = 52,
    width=9cm,
    enlargelimits=0.3,% these limits are used to shrink or expand the graph. The lesser the limit, the higher the graph will expand or grow. The greater the limit, the more graph will shrink.   
    % legend style={at={(1.3,1.0)}, % these are the measures of the bottom row containing surplus (wheat, Tea, rice), where -0.25 is the gap between the bottom row and the graph.   
    %   anchor=north,legend columns=-1},
    legend style={at={(1.52,1.0)}},
    legend style={cells={align=left}},
    legend style={nodes={scale=0.7, transform shape}},
    legend image code/.code={
        \draw [#1] (0cm,-0.1cm) rectangle (0.2cm,0.25cm); },
      % here, north is the position of the bottom legend row. You can specify the east, west, or south direction to shift the location.   
    ylabel={\ CM-Score}, % there should be no line gap between the rows here. Otherwise, latex will show an error.  
    xlabel={\ Number of Leads},
    symbolic x coords={12-Leads, 3-Leads, 2-Leads},  
    xtick=data,  
    nodes near coords={%
                    \pgfmathprintnumber\pgfplotspointmeta\%%
                },
                % set the style of `nodes near coords'
    every node near coord/.append style={
        font=\scriptsize,
        color=black,
        /pgf/number format/fixed,
    },
    nodes near coords style={rotate=90,font=\tiny} ,  
    nodes near coords align={horizontal} ,  
    ]  
\addplot[fill=gray!30] coordinates {(12-Leads,60.46) (3-Leads, 58.49) (2-Leads, 56.66)}; % these are the measures of a particular bar graph. The tick marks of the y-axis will be adjusted automatically according to the data values entered in the coordinates.  
\addplot[fill = blue!70] coordinates {(12-Leads,60.67) (3-Leads, 59.49) (2-Leads, 57.27)};  
\addplot[fill = red!70] coordinates { (3-Leads, 60.39) (2-Leads, 58.43)};  
\legend{LRH-Net,LRH-Net+Vanilla-KD,LRH-Net+MLKD (Vanilla-KD)}  
\end{axis}  
\end{tikzpicture}  
\caption{Bar-Plot showing the improvement in LRH-Net's CM-score using Vanilla knowledge distillation (KD) and Multi-level Knowledge Distillation using Vanilla-KD.} \label{bar_improvement}
\end{figure}

Knowledge distillation helps LRH-Net to further reduce its size or parameters with the reduction in number of leads which causes a steady decrease in performance too. This makes it easier to utilise on an edge device because only a small number of electrodes are needed to capture the requisite lead data. In MLKD, KL losses for the outputs of the student network with both teacher networks result in some amount of conflict in the gradients between these two losses. In the sequential configuration, the second teacher network has been trained to reproduce the outputs of the first teacher network. This will result in less conflict between the two KL loss terms, as the outputs will be closer in the latent-space. In the case of the parallel configuration, both KD loss terms will be in conflict with each other simultaneously to make the outputs of the student network close to each of the teacher networks.

The experimental results shown in the Tables (\ref{tab3}~\&~\ref{tab4}) show that MLKD not only generalizes LRH-Net on fewer leads but also makes it more robust and accurate. LRH-Net after distillation using MLKD with Vanilla KD on 3-Leads is performing almost similar to LRH-Net on 12-Leads (Fig~\ref{bar_improvement}). This could be because the V2 lead (calculated using chest electrode) in a 3-lead configuration has more disease-specific information. However, the low-parameter model is drastically affected in the F1 scores of few hard-to-classify diseases like Sinus Arrhythmia (SA). 

\section{Conclusion}
In this work, we propose the Low Resource Heart-Network (LRH-Net) for detecting cardiovascular diseases. The proposed model is evaluated on a combination of four large heterogenous datasets provided by the PhysioNet-2020 challenge. The proposed low resource model not only enables edge computing on a wearable device but also gives better results as compared to other architectures proposed for wearable devices previously. In addition to compressing the model, our work also focused on using reduced number of leads to reduce the input processing without sacrificing much on the performance using multi-level knowledge distillation (MLKD). As a result, the computational resources, cost, and input channels required for our proposed model are reduced. This approach is carefully designed to increase the ease of use and affordability of an accurate edge device in rural or semi-urban areas.  Further research can focus on lowering the performance gap in low-lead configurations by optimizing the number of steps required to distill majority of the critical information by varying the levels of MLKD so that the classification performance on hard-to-classify diseases does not get severely affected.

%\subsubsection{Acknowledgements} ***** Anonymous .

\bibliographystyle{splncs04}
\bibliography{9}

\end{document}